\documentclass[%
 reprint,amsmath,amssymb,aps]{revtex4-2}

\usepackage{graphicx}
\usepackage{dcolumn}
\usepackage[export]{adjustbox}
\usepackage{bm}
\usepackage{tabularx,multirow,array,booktabs,makecell,float}
\usepackage[utf8]{inputenc}
\usepackage[T1]{fontenc}
\usepackage[abs]{overpic}
\usepackage{xcolor}
\begin{document}
\title{Frequency subspace encoding for multiplexed quantum secret sharing
}

\author{Meritxell Cabrejo-Ponce}
\email{meritxell.cabrejo.ponce@iof.fraunhofer.de}
\author{Carlos Sevilla-Gutiérrez}
\affiliation{Friedrich Schiller University Jena, Institute of Applied Physics, Abbe Center of Photonics, Albert-Einstein-Str. 15, 07745 Jena, Germany}%
\affiliation{Fraunhofer Institute for Applied Optics and Precision Engineering, Albert-Einstein-Strasse 7, 07745 Jena}
\author{Christopher Spiess}
\affiliation{Fraunhofer Institute for Applied Optics and Precision Engineering, Albert-Einstein-Strasse 7, 07745 Jena}
\author{Fabian Steinlechner}
\email{fabian.steinlechner@iof.fraunhofer.de}
\affiliation{Friedrich Schiller University Jena, Institute of Applied Physics, Abbe Center of Photonics, Albert-Einstein-Str. 15, 07745 Jena, Germany}%
\affiliation{Fraunhofer Institute for Applied Optics and Precision Engineering, Albert-Einstein-Strasse 7, 07745 Jena
}%


\begin{abstract}
Quantum secret sharing (QSS) is a multi-party quantum communication protocol that can be realized with bipartite entanglement and relative phase encoding. 
Previous implementations typically encoded the phase in the pump, applying it across the entire source bandwidth, thereby limiting scalability via wavelength multiplexing. 
In contrast, we present a variant of the standard QSS protocol that leverages frequency correlations to connect multiple users with a single source. 
The secret owner, who has access to the source, encodes classical information by applying frequency-dependent phase modulation to a broadband polarization-entangled photon pair. Each frequency channel therefore provides an independent QSS session among the secret owner and a pair of users. 
We demonstrate state fidelities of at least 90\,\% for a channel pair of the 200\,GHz ITU grid, 
which could be extended to more than 40 frequency bins with adequate dense-wavelength division multiplexed filters. 
Our results provide a resource-efficient path toward multi-user secret sharing over wavelength-multiplexed networks, eliminating the need for multiple two-photon or multi-photon sources.
\end{abstract}

\maketitle

\section{Introduction}
Quantum communication aims to securely distribute quantum or classical information between various parties, often by means of entanglement. Quantum key distribution (QKD) is a prominent application in this field \cite{pirandola_advances_2020}. It benefits from the random outcomes of the measurement of quantum states to generate encryption keys that are secretly shared between a pair of users \cite{bennett_quantum_1992}. As experimental demonstrations move out of the lab into field experiments \cite{neumann_continuous_2022,pelet_operational_2023, krzic_towards_2023}, efforts also focus on the scalability of quantum networks as well as protocols beyond QKD. Quantum networks interconnect multiple users simultaneously and enable various applications such as bipartite key exchange 
between any two parties in the network. This may be achieved 
with wavelength multiplexing techniques and a single quantum source \cite{wengerowsky_entanglement-based_2018,joshi_trusted_2020}, 
facilitating optimal bandwidth allocation to maximize key rates for every channel \cite{appas_flexible_2021,lingaraju_adaptive_2021}. In contrast, protocols beyond QKD often focus on cryptographic primitives that involve more than two parties, thereby unlocking the potential of highly connected quantum networks. 
Among these, 
Quantum Secret Sharing (QSS) 
is a protocol that splits a secret among several parties who must collaborate to jointly reconstruct it \cite{cleve_how_1999,hillery_quantum_1999}. 

Often, protocols beyond QKD, including QSS, rely on multipartite entanglement by sharing Greenberger-Horne-Zeilinger (GHZ) multi-photon states \cite{chen_multi-partite_2007, murta_quantum_2020}.
Yet, multi-photon entanglement is challenging, usually limited to low generation rates \cite{proietti_experimental_2021, meyer-scott_scalable_2022, pickston_conference_2023}. 
More resource efficient approaches have also been proposed, one of which relies on the sequential optical modulation of a single qubit \cite{schmid_experimental_2005,bogdanski_experimental_2008, scherpelz_entanglement-secured_2011, hai-qiang_experimental_2013} or qudit \cite{yu_quantum_2008, smania_experimental_2016, pinnell_experimental_2020} by each party. 
This approach favors user scalability at the expense of introducing additional loss with multiple 
high-speed modulation devices, 
thereby reducing the experimental photon rates. Alternatively, sets of two-photon entangled states can be used to encode information and perform QSS for $N_\text{QSS}=3$ parties \cite{karlsson_quantum_1999}. For example, practical sets of states are those that encode information in their relative phase 
\cite{tittel_experimental_2001}. Whereas high-quality two-photon entanglement is readily obtained at high rates and already serves as a cornerstone of many QKD experiments \cite{neumann_continuous_2022,pelet_operational_2023, krzic_towards_2023}. 
This approach has been investigated with manual modulation for time \cite{tittel_experimental_2001} and polarization qubits \cite{williams_quantum_2019}, and recently extended to practical near-GHz polarization modulation \cite{cabrejo-ponce_ghz-pulsed_2022}. 
For protocols involving more than three parties ($N_\text{QSS}>3$), one can allow additional parties to modulate the passing photon \cite{williams_quantum_2019}. These methods, however, encode a phase over the entire bandwidth of the source, thus limiting the implementation of the protocol in wavelength-multiplexed quantum networks.

Here we demonstrate that a single source of bipartite entanglement can also be used to extend QSS to multi-user networks. 
Our approach supports wavelength-multiplexed scalability by addressing each wavelength channel pair or subspace independently. To implement them, we employ simple spectral shaping tools that are fully compatible with the telecom infrastructure. 
We achieve state fidelities $>\,90\,\%$ for the selected QSS channels 13 and 60 within the 200\,GHz ITU grid. These measurements are performed while simultaneously monitoring the visibility for a single state in the ITU channels 16 and 57, demonstrating the viability of subspace encoding.  
The scheme introduced here enables independent and parallel realization of polarization-encoded QSS between the dealer, the secret owner, 
and multiple pairs of users, while requiring a single source at the center of the network.

\section{Quantum Secret Sharing}
In QSS, the dealer, Charlie, splits a secret between two parties, Alice and Bob, ensuring that neither of them has complete information. Therefore, Alice and Bob must collaborate to reconstruct the information sent by Charlie. 
In our implementation, the dealer encodes classical information in a two-photon entangled state \cite{karlsson_quantum_1999}. For the information to remain secret, the encoding must utilize sets of states that are mutually unbiased. This way, no measurement can distinguish all states simultaneously due to the no-cloning theorem \cite{nielsen_quantum_2010}. 
For convenience, we choose to encode the information in the phase of a maximally entangled state of the form \cite{tittel_experimental_2001,williams_quantum_2019}: 
\begin{equation}
|\psi(\theta)\rangle = \frac{1}{\sqrt{2}} \Big( 
|+z,+z\rangle + e^{i\theta} |-z,-z\rangle
\Big)
\end{equation}
where the $|\pm z\rangle$ states denote the eigenstates of the $Z$ basis, the logical basis of the qubit (e.\,g. a polarization qubit is typically expressed as $|+z\rangle\equiv|H\rangle$ and $|-z\rangle\equiv|V\rangle$). The phase is randomly chosen from the set $\theta\in\{0,\pi,\pi/2,3\pi/2\}$, corresponding to two sets of states that are mutually unbiased. 
For simplification, these states are referred as $|\phi^\pm\rangle$ for $\theta=\{0,\pi\}$, and $|\varphi^\pm\rangle$ for $\theta=\{\pi/2,3\pi/2\}$. Hence, the states in each set are orthogonal and can be used to encode distinct bits. 
The maximally entangled Bell states $|\phi^\pm\rangle$ display maximal correlations in the joint $X\otimes X$ and $Y\otimes Y$ bases:
\begin{align}\label{eq:phi_p}
|\phi^+\rangle 
& =
\frac{1}{\sqrt{2}} \Big( 
|+z,+z\rangle + |-z,-z\rangle
\Big)
\\ \nonumber 
& =
\frac{1}{\sqrt{2}} \Big( 
|+x,+x\rangle + |-x,-x\rangle
\Big)
\\ \nonumber 
& =
\frac{1}{\sqrt{2}} \Big( 
|+y,-y\rangle + |-y,+y\rangle
\Big)
\end{align}

\begin{align}\label{eq:phi_m}
|\phi^-\rangle 
& =
\frac{1}{\sqrt{2}} \Big( 
|+z,+z\rangle - |-z,-z\rangle
\Big)
\\ \nonumber 
& =
\frac{1}{\sqrt{2}} \Big( 
|+x,-x\rangle + |-x,+x\rangle
\Big)
\\ \nonumber 
& =
\frac{1}{\sqrt{2}} \Big( 
|+y,+y\rangle + |-y,-y\rangle
\Big)
\end{align}
with $|\pm x\rangle$ and $|\pm y\rangle$ denoting the eigenvectors of the $X$ and $Y$ bases, respectively. These bases correspond to the diagonal and circular bases of a polarization qubit. 
On the other hand, the states $|\varphi^\pm\rangle$ are also maximally entangled, despite not being the commonly used Bell states. 
They display maximal correlations for different basis combinations. Namely, for $X\otimes Y$ and $Y\otimes X$:
\begin{align}\label{eq:varphi_p}
|\varphi^+\rangle 
& =
\frac{1}{\sqrt{2}} \Big( 
|+z,+z\rangle + i |-z,-z\rangle
\Big)
\\ \nonumber 
& =
\frac{1}{\sqrt{2}} \Big( 
|+x,+y\rangle + |-x,-y\rangle
\Big)
\\ \nonumber 
& =
\frac{1}{\sqrt{2}} \Big( 
|+y,+x\rangle + |-y,-x\rangle
\Big)
\end{align}
\begin{align}\label{eq:varphi_m}
|\varphi^-\rangle 
& =
\frac{1}{\sqrt{2}} \Big( 
|+z,+z\rangle -i |-z,-z\rangle
\Big)
\\ \nonumber 
& =
\frac{1}{\sqrt{2}} \Big( 
|+x,-y\rangle + |-x,+y\rangle
\Big)
\\ \nonumber 
& =
\frac{1}{\sqrt{2}} \Big( 
|-y,+x\rangle + |+y,-x\rangle
\Big)
\end{align}
To identify the encoded state, each user needs to measure their photon either in the $X$ or $Y$ basis and then compare their measurement result with the other participants in the protocol. 
Neither of them alone can retrieve the actual phase information, given that each measurement outcome can be related to all four states. 
Therefore, each party in the protocol needs to make public the chosen basis for encoding and decoding. The dealer shares the encoding basis $C\in\{\phi,\varphi\}$, whereas Alice and Bob communicate the measurement basis $A,B\in\{X,Y\}$. The private bit of Charlie, the secret, corresponds to the sign of the state in the encoding basis $c\in\{+,-\}$, and the private bits of the other two users result from the sign of the measurement outcome, $a,b\in\{+,-\}$. 

As in standard QKD, basis reconciliation is also necessary in this protocol. Only four combinations of public bits yield meaningful results, 
hence the other combinations need to be discarded. 
Furthermore, equations (\ref{eq:phi_m}-\ref{eq:varphi_m}) 
show that most of the encoding and measurement combinations display positive correlated outcomes in the $X$ and $Y$ measurement bases. 
That is, the parity of private bits $\epsilon=abc$ is positive most of the time, $\epsilon=+1$, 
except for the public bit combination $(A,B,C)=(Y,Y,\phi)$. 
In this case, the users simply need to flip their secret bit to be in agreement with Charlie's bit.
To guarantee the security of the protocol, the encoding as well as the measurement bases should be chosen at random. This allows the participants to track the quantum bit error rate (QBER), not only to detect a possible eavesdropper, but also to test, in pairs, whether the other participant may be dishonest \cite{hillery_quantum_1999,karlsson_quantum_1999,williams_quantum_2019}. 
Finally, standard error correction protocols may be implemented. 

\section{Frequency-subspace QSS}
\begin{figure}
    \centering
    \includegraphics[width=0.975\linewidth]{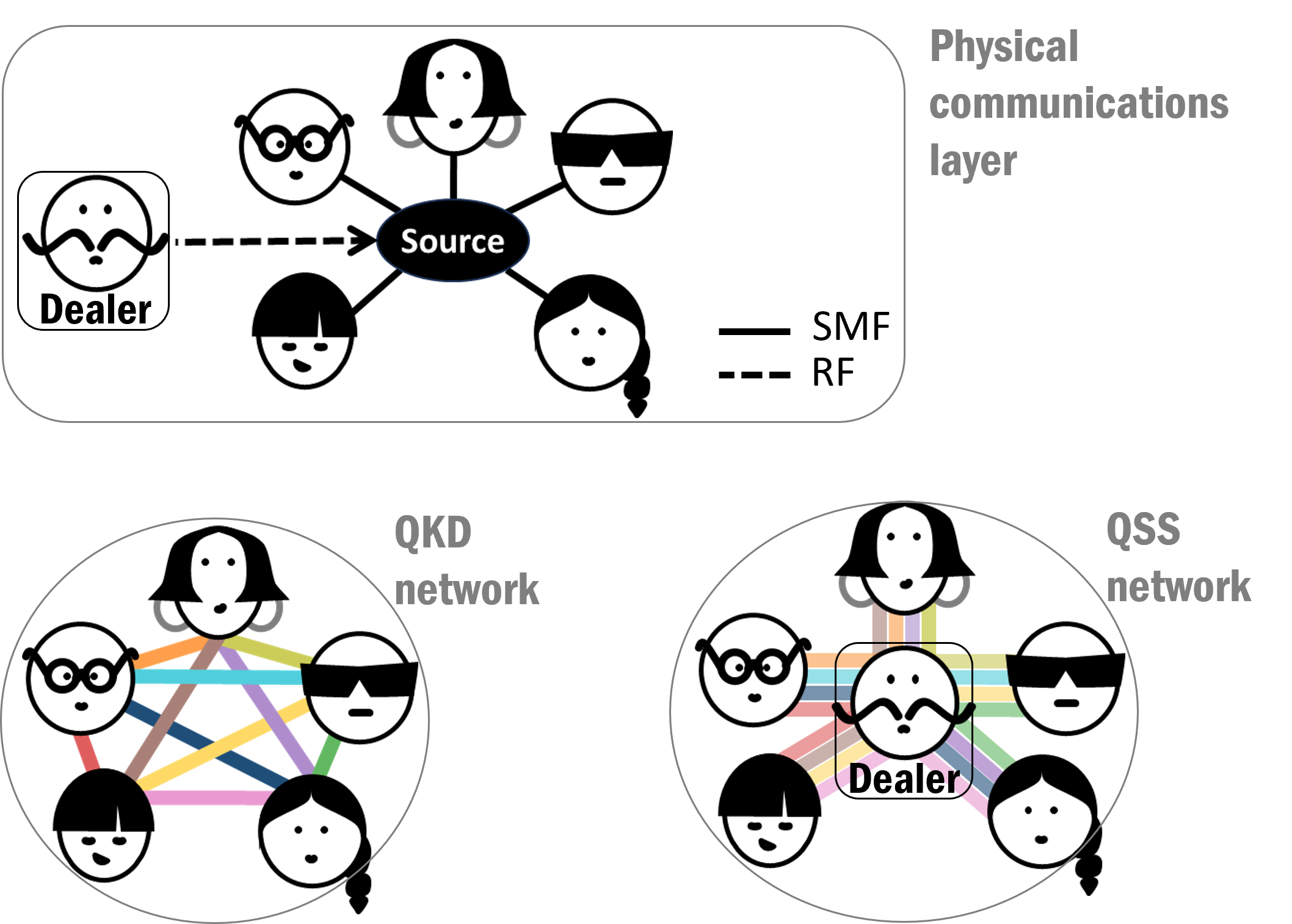}
    \caption{Top: Representation of the physical connection layout of a quantum network, here with five users in a star topology. All users of the network are connected to the photon-pair source via a single mode fiber (SMF). The dealer in quantum secret sharing (QSS) has access to the source and introduces a relative phase with electrical radio-frequency (RF) signals. Bottom left: a standard fully-connected quantum network using the same physical layout for quantum key distribution (QKD). Here, each user pair shares an entangled state according to the frequency correlations of the quantum source. 
    Bottom right: a fully-connected network for QSS requires the dealer to actively encode information independently in each channel pair of the shared entangled state.}
    \label{fig:network}
\end{figure}

The relative phase of an entangled state is determined by the properties of the pump driving the process. 
Therefore, previous demonstrations of QSS employed manual \cite{tittel_experimental_2001,williams_quantum_2019} and electro-optic modulation \cite{cabrejo-ponce_ghz-pulsed_2022} techniques to control the relative phase of the pump state.  
However, this phase is applied uniformly across the entire source bandwidth, making it incompatible with standard QKD networks that utilize wavelength division multiplexing (WDM) schemes, as displayed in figure \ref{fig:network} (top). 
These networks are advantageous because they allow multiple wavelength channels to be transmitted simultaneously over a single optical fiber to each user. This architecture reduces infrastructure costs and enables more flexible, dynamic network configurations.  
In the quantum layer, 
entanglement is distributed from a single quantum source among multiple user pairs by assigning each pair a correlated frequency channel, as represented by the coloured connections in figure \ref{fig:network} (bottom left). 
If QSS were applied directly in this configuration, every channel pair assigned to different users would carry the same encoding. 
In contrast, an independent implementation of QSS requires each channel to be encoded separately. 
With this idea in mind, figure \ref{fig:network} (bottom right) shows the dealer, who might be representing the bank or the service provider, located at the center of the network. They encode the phase independently for every communication channel and performs QSS with various pairs of users. To achieve this, we propose addressing the frequency subspaces of a broadband polarization-entangled photon source.

\begin{figure}[ht]
\centering
\includegraphics[width=0.49\textwidth]{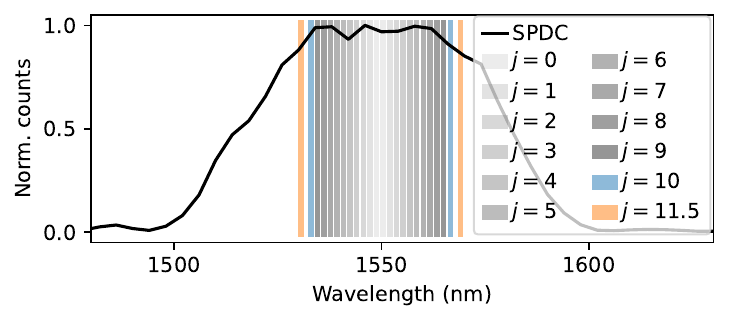}
\caption{Sinc-shaped spectrum of our broadband photon pair source, based on spontaneous parametric down-converson (SPDC). The spectrum is discretized into multiple frequency channel pairs $j$, symmetric with respect to the degeneracy wavelength $\lambda_0=1548.1\,$nm. Due to filter availability, we use channel pair $j=10$ (blue) for interferometer stabilization, and channel pair $j=11.5$ (orange) (i.\,e. a 200\,GHz bandwidth channel between $j=11$ and $j=12$) for implementing subspace-encoded QSS.}
\label{fig:spectrum}
\end{figure}

Consider a standard polarization-entangled photon-pair source based on spontaneous parametric processes and pumped with a continuous-wave laser. 
Due to energy conservation, these processes naturally provide 
frequency anti-correlation for the signal and the idler photon. 
Discretization of the spectrum yields polarization-entangled states only for correlated frequency channels. 
Every pair $j$ of correlated channels is described by $|j,j\rangle_{s,i} = 
|\omega_0+j\Delta\omega, \omega_0-j\Delta\omega\rangle$, with $\Delta\omega$ the free spectral range (FSR) and $\omega_0$ the degeneracy frequency, as shown in figure \ref{fig:spectrum}.
The overall state is expressed as:
\begin{equation}\label{eq:state_hyper}
    |\psi \rangle = \frac{1}{\sqrt{2}} \Big(
    |HH\rangle + |VV\rangle 
    \Big)\otimes \frac{1}{\sqrt{d}} \sum_{j=1}^d|j, j\rangle  
\end{equation}
which is hyper-entangled in the polarization and frequency DOFs, although the frequency entanglement is simply disregarded. By manipulating the relative spectral phase $\theta_j$ between the two 
polarization components ($|H\rangle$ and $|V\rangle$) of either or both photons, independently for every mode pair $j$, the subspace encoding is achieved:
\begin{equation}\label{eq:state_j}
    |\psi (\theta_j)\rangle_j = \frac{1}{\sqrt{2}} \Big(
    |HH\rangle\otimes|j,j\rangle + e^{i\theta_j} |VV\rangle \otimes |j,j\rangle
    \Big) 
\end{equation}
This can be accomplished with a balanced polarization-dependent Mach-Zehnder interferometer (MZI), that distinguishes between the two polarization components, and a pulse shaper. The latter device is used to set the relative phase $\theta_j$ for each frequency channel $|j,j\rangle$ independently. 

\begin{figure*}[ht]
\centering
\includegraphics[width=0.85\textwidth]{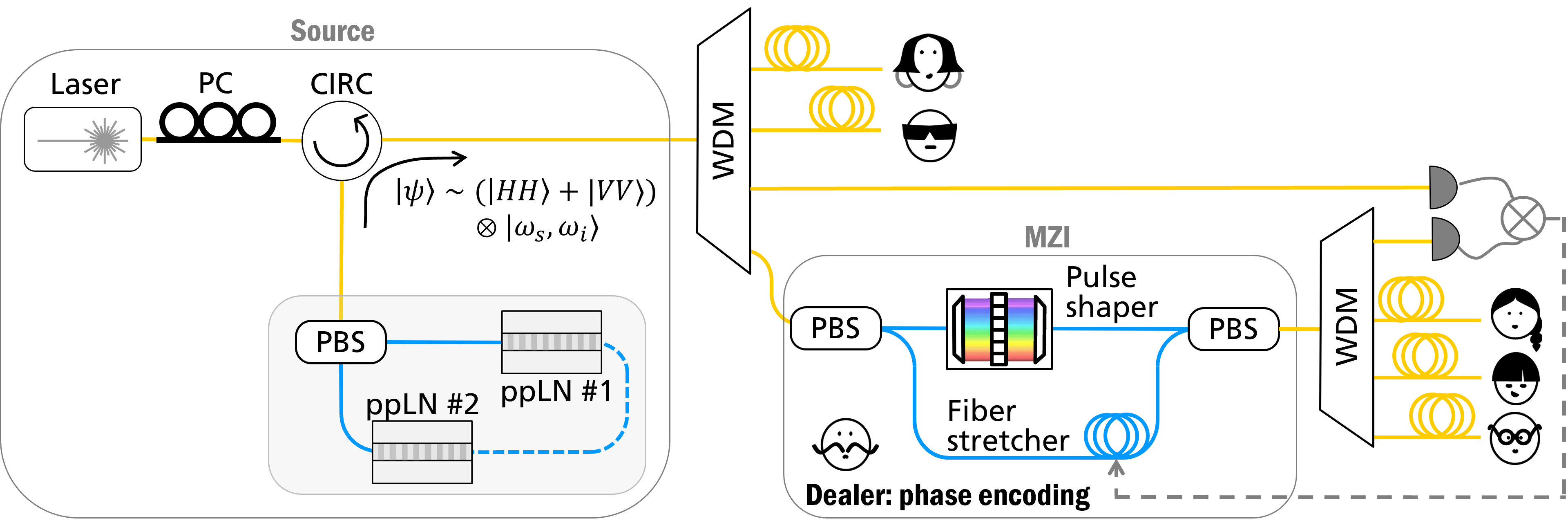}
\caption{Setup of the reconfigurable all-fiber polarization-entangled photon source. Its broadband spectrum is sliced into multiple
channels with a wavelength division multiplexer (WDM). One of the photons of each pair is sent to a polarization-dependent Mach-Zehnder interferometer (MZI) to control the relative spectral
phase of the two polarization components. At the output, the photon is distributed
into the network via a second WDM. The MZI is stabilized with the
quantum bit error rate of a pair of correlated frequency channels. The other elemenents in the figure are: circulator (CIRC), polarization beam splitter  (PBS), polarization controller (PC), periodically-poled Lithium niobate waveguide (ppLN).}
\label{fig:Setup}
\end{figure*}

\section{Setup}
To demonstrate this concept, we use the polarization entangled photon source introduced in \cite{cabrejo-ponce_ghz-pulsed_2022}, as shown in figure \ref{fig:Setup}. The source consists of cascaded second harmonic generation (SHG) and spontaneous parametric down-converson (SPDC) processes in two distinct fiber-coupled periodically poled lithium niobate waveguides (ppLN) from Covesion. They are integrated into an all-fiber Sagnac interferometer, which naturally provides phase stability. The advantage of this configuration is that it uses only telecom and commercial off-the-shelf components (COTS) as well as the natural filtering provided by shorter-wavelength fibers to remove the remaining pump. 
The generated photon pairs are centered at $\lambda_0=2\pi c/\omega_0=$1548.1\,nm (193.65\,THz), at the intersection of the International Telecommunication Union (ITU) channels 36 and 37 of the 100\,GHz grid in G.694.1, and extend over 70 nm of bandwidth. After photon pair generation, each pair is separated via wavelength division multiplexers (WDMs) prior to distribution. 
The polarization-entangled photons achieve fidelities of 99\,\% for the $|\phi^+\rangle$ state and for different WDM bandwidths, as already shown in previous work from 1 to 14 nm \cite{cabrejo-ponce_ghz-pulsed_2022, ancsin_modulated_2023}. 
In the present demonstration, the signal photon, the higher energetic photon, is directly transmitted to the user. The idler photon is first spectrally manipulated by the dealer, to encode the phase of the protocol, before distribution.

The dealer uses an MZI to separate the two polarization components for independent processing. 
One arm incorporates a commercially available and fiber-coupled pulse shaper (Waveshaper 16000), which is polarization independent, to control the relative spectral phase. 
In our experimental implementation, this arm contains approximately 10\,m of optical fiber, resulting in an interferometer that is highly sensitive to external vibrations and temperature fluctuations. 
To compensate this, the other arm includes a variable optical delay line, to carefully match the path length, and a fiber stretcher to actively stabilize the relative phase between the two arms. 
The feedback signal for phase stabilization is obtained from the quantum bit error rate (QBER) of an unused pair of frequency channels (the middle channels in figure \ref{fig:Setup}) \cite{spiess_stabilization_2023}. In our demonstration, we use 200\,GHz ITU channels 16 and 57, namely $j=10$ in equation (\ref{eq:state_hyper}) and figure \ref{fig:spectrum}. 
In particular, the QBER is tracked in the $X$ basis with a multi-outcome polarization analysis module, i.\,e. with projection measurements $\{|x,x\rangle, |x,-x\rangle, |-x,x\rangle, |-x,-x\rangle\}$. This allows us to obtain an error signal that is independent of intensity fluctuations from the source and does not require the use of external lasers \cite{spiess_stabilization_2023}.

For this proof-of-concept experiment, the QBER is minimized by using a dither signal that is 10\,\% of the stretcher’s $V_\pi$, with a feedback time of 20\,Hz.  
Therefore, even without phase fluctuations in the MZI, our algorithm checks whether a different bias point provides a lower QBER, reducing the average visibility.
While the unstabilized MZI displays visibilities of 99\,\% over short time scales, the necessary stabilization reduces it to 80-90\,\% on average, depending on the strength of the phase fluctuations. Note, however, that the active stabilization is crucial for our implementation of QSS.
An unstable interferometer changes the relative phase of the entangled state by $\theta_\text{MZI}$, which rapidly changes over time. 
Thus, the state in equation (\ref{eq:state_j}) now depends on both the established phase by Charlie $\theta_j$ and the MZI phase, $|\psi (\theta_j+\theta_\text{MZI})\rangle_j$.  
This limitation, nevertheless, can readily be circumvented either by using shorter fibers and passive stabilization, active stabilization, or a polarization-sensitive spectral phase modulator in an interferometric-stable configuration.

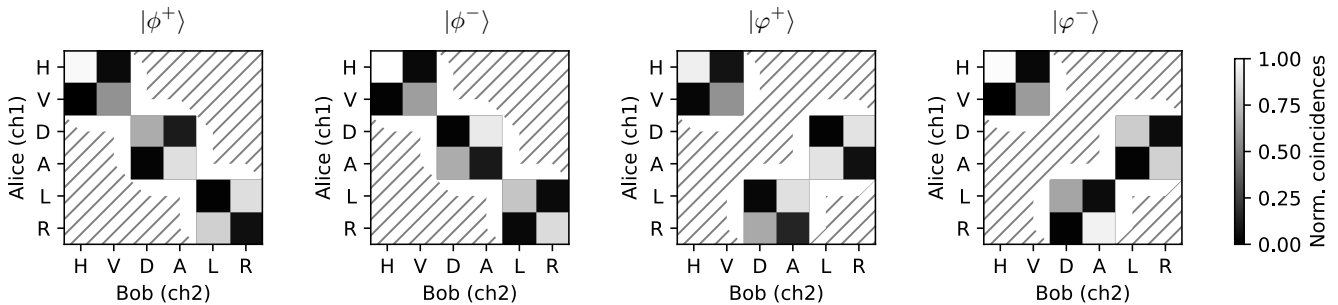
\begin{figure*}
\centering
    \begin{overpic}[width=\linewidth]{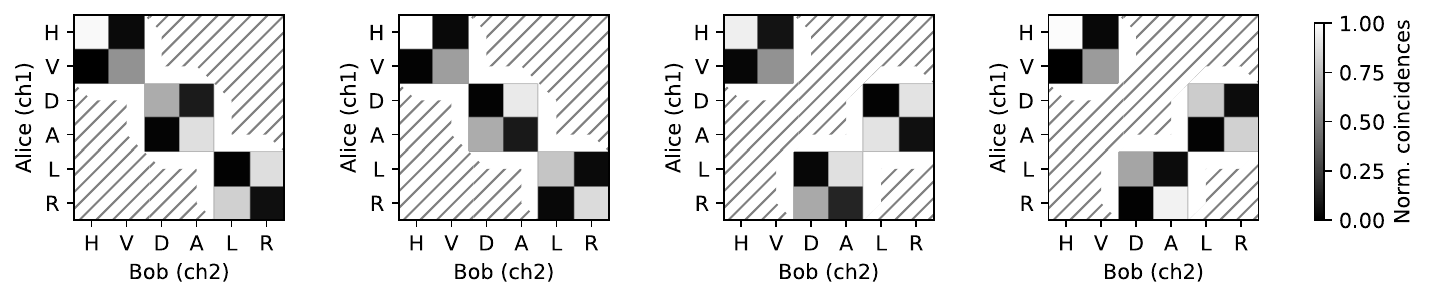}
        \put(55,110){$|\phi^+\rangle$}
        \put(170,110){$|\phi^-\rangle$}
        \put(285,110){$|\varphi^+\rangle$}
        \put(400,110){$|\varphi^-\rangle$}
    \end{overpic}
\caption{
Experimental correlations of the four QSS states in our multiplexed network. These results are obtained with the channel pair 13-60 of the 200\,GHz ITU grid and follow the expected correlations from equations (\ref{eq:phi_p}-\ref{eq:varphi_m}). 
$\{H,V\}$, $\{D,A\}$ and $\{R,L\}$ are the eigenstates of the $Z$, $X$ and $Y$ bases, respectively. The hashed areas have not been measured. The resulting visibilities are reported in table \ref{tab:results}.
}
\label{fig:Tomo}
\end{figure*}

\section{Results}
After setting $\theta_\text{MZI}=0$, we can proceed with the phase encoding and polarization analysis. Due to filter availability, we use the 200\,GHz ITU channels 60 and 13, corresponding to channel pairs $j=11.5$ according to the spectral grid in figure \ref{fig:spectrum}. The spectral phase is chosen from the set $\theta_j\in\{0,\pi,\pi/2,3\pi/2\}$ and is applied to channel 60. 
Each phase setting is individually characterized by deriving metrics of the produced state. 
To reduce the number of measurement settings required for full state tomography, and therefore the measurement time, only the correlated bases are measured, as displayed in figure \ref{fig:Tomo}. This is sufficient to compute the fidelity and the correlation parameter $\langle\epsilon\rangle$. 
The fidelity describes how close our experimental state $\rho_\text{exp}$ is with respect to the ideal and pure target state $\rho$, and is computed as $F(\rho_\text{exp},\rho)=\text{Tr}\{\rho_\text{exp}\rho\}$. 
It can also be expressed in terms of the visibilities of the correlated bases \cite{jennewein_performing_2009}, 
$F(\rho_\text{exp},|\phi^\pm\rangle\langle\phi^\pm|) = \frac{1}{4}(1\pm V_{XX}\mp V_{YY} + V_{ZZ})$ and $F(\rho_\text{exp},|\varphi^\pm\rangle\langle\varphi^\pm|) = \frac{1}{4}(1\mp V_{XY}\mp V_{YX} + V_{ZZ})$, where $V_{AB}$ correspond to the joint measurement with bases $A$ and $B$.
Each visibility is computed with four measurement settings:
\begin{equation}
    V_{AB}=\frac{ R_{mm} - R_{mn} - R_{nm} + R_{nn} }
    { R_{mm} + R_{mn} + R_{nm} + R_{nn} }
\end{equation}
where $R_{mn}$ corresponds to the number of coincidence events with measurement settings $|m\rangle$, the eigenvectors of basis $A$, and $|n\rangle$, the eigenvectors of basis $B$. 
The experimental fidelities are shown in table \ref{tab:results}.

The correlation parameter, on the other hand, describes how well each orthogonal state ($|\phi^\pm\rangle$ or $|\varphi^\pm\rangle$) can be identified in a joint basis, thus allowing the private bit $c=\pm$ to be recovered. 
Its expectation value can be computed for every combination of public bits with the average visibility of the orthogonal states 
\cite{williams_quantum_2019}. 
Let us denote $V_{c}$ 
the visibility of each state, obtained with the joint measurement bases $A\otimes B$, 
according to the public bit $C$ and private bit $c$ of Charlie. 
Then, the correlation parameter is obtained as $\langle\epsilon\rangle = \frac{1}{2}\big( V_{c=+} - V_{c=-}\big)$, assuming that Charlie chooses his private bit with equal probability \cite{williams_quantum_2019}. The experimental results $\langle\epsilon\rangle_\text{exp}$ are displayed in Table \ref{tab:results}.

The obtained correlation parameters 
are lower than the reported fidelities. This is because $\langle\epsilon\rangle_\text{exp}$ only depends on the visibilities in the $X$ and $Y$ bases, which are highly influenced by the noise of the MZI stabilization. In contrast, the fidelity includes information of the $Z$ basis, which is not affected by phase fluctuations. 
Nevertheless, as long as the QBER is below 11\,\% assuming unity correction efficiency, the key fraction in the asymptotic limit is positive \cite{koashi_secure_2003}. To compare this, the QBER can be estimated as $\frac{1}{2}(1-|\langle \epsilon\rangle|)$ \cite{gisin_quantum_2002,williams_quantum_2019}. The asymptotic key fraction per coincidence detection can then be obtained with the standard Koashi-Preskill formula, $r_\infty = 1-2h( \text{QBER} )$ \cite{koashi_secure_2003}, where $h(x)=-x\log_2(x)-(1-x)\log_2(1-x)$ is the binary entropy function. 
The results are also presented in the same table. The imperfect visibilities reduce the correlation parameter to $\langle |\epsilon_\text{exp}|\rangle=0.894$ on average. This translates to an average QBER of 5.3\,\% and a key fraction of 0.4\,bits/photon.

\begin{table}[h!]
    \centering
    \begin{tabular}{c|c c c c}
    \toprule
        State & $|\phi^+\rangle$ & $|\phi^-\rangle$ & $|\varphi^+\rangle$ & $|\varphi^-\rangle$ \\ \midrule
        Fidelity (\%) & 93.0 & 92.9 & 90.3 
        & 95.4 \\ \midrule
        $V_{XX}$ & 85.3  & -86.4& - & - \\
        $V_{YY}$ & -92.3 & 91.3 & - & - \\
        $V_{XY}$ & -     & -    & 93.0 & -91.7 \\
        $V_{YX}$ & -     & -    & 92.9 & -82.6 \\
        \bottomrule
        \toprule
        Public bits $(A,B,C)$ & $(X,X,\phi)$ & $(Y,Y,\phi)$ & $(X,Y,\varphi)$ & $(Y,X,\varphi)$ \\ \midrule
        $\epsilon_\text{ideal}$  & 1 & -1 & 1 & 1 \\
        $\langle \epsilon \rangle_\text{exp}$  & 0.86 & -0.92 & 0.92 & 0.88 \\
        QBER (\%)  & 7.1 & 4.1 & 3.8 & 6.1 \\
        $r_\infty$ (bits/photon) & 0.26 & 0.51 & 0.53 & 0.34 \\
    \bottomrule
    \end{tabular}
    \caption{
    Fidelities and visibilities $V_{AB}$ from the prepared entangled states. We also compute the correlation parameter $\langle \epsilon \rangle$, QBER and asymptotic key fraction $r_\infty$ for each public bit combination with meaningful results. 
    }
    \label{tab:results}
\end{table}

\section{Discussion}
Our source provides a bandwidth of 70 nm that is aligned to the ITU frequency grid. This enables the discretization of its spectrum into multiple channels, according to the spectral filters of choice. 
The most standard telecom grid uses 100\,GHz spacing, providing at least 44 
pairs of frequency channels with our source. 
In the present example, we have used the 200\,GHz grid due to filter availability, which offers 22 
channel pairs. 
A number of $N_\text{MUX}(N_\text{MUX}-1)/2$ channel pairs enables the full connectivity of $N_\text{MUX}$ users \cite{wengerowsky_entanglement-based_2018}. Therefore, our source enables fully-connected networks of up to 9 users with DWDMs of the 100\,GHz ITU grid and up to 7 users with the 200\,GHz ITU grid. 
In addition, active routers allow the reconfigurability of allocated spectral channels and their bandwidth. This capability facilitates dynamic user connections and the optimization of key rates based on link conditions \cite{appas_flexible_2021,lingaraju_adaptive_2021}.
While our source meets the QKD requirements in these networks, our proposed scheme further enhances the network services by also offering the possibility of QSS.  
Although the dealer is positioned at the center of the network to implement QSS with multiple user pairs, other users with access to spectral phase control may also actively participate in the protocol.

In our work, we have demonstrated spectral phase modulation for a single frequency subspace, independent of the frequency channels used for active locking of the interferometer. 
This indicates that spectral manipulation can be extended to the remaining multiplexed channels. 
It should be noted that our interferometer has been stabilized only with the feedback of the quantum signal \cite{spiess_stabilization_2023}, which consumes one pair of channels but avoids the use of external lasers. 
Nevertheless, there are several aspects that can be optimized. 
For instance, the locking scheme can be improved by including passive stabilization. Reducing the impact of external noise (thermal or mechanical), allows to decrease the dither signal and results in higher visibilities. 
Moreover, the interferometer and available filters introduced an insertion loss of $\sim\,$7\,dB for the signal photon and $\sim\,$10\,dB for the idler photon, which can be lowered using custom components. 
Simple improvements, such as an optimized active stabilization that achieves 98\,\% visibility would double the QSS key fraction at the same photon rate (more details in the appendix). That is, the key fraction would directly increase from 0.4 to 0.8\,bits/photon in our experiment.
Whereas a loss reduction of 7\,dB for each photon would increase $\times5$ the heralding efficiency and $\times50$ the total key rate. 

On the other hand, it is also highly relevant to increase the modulation speed. The commercially-available pulse shaper used in our experiment only achieves Hz rates.
Nevertheless, it is expected that future technological advances, such as on-chip interferometers with inherent phase stability, will eliminate the need of 
active stabilization. Newer designs for on-chip modulation already enable wavelength-parallelized encoding exceeding GHz-rates 
\cite{liu_first_2024}. 
Furthermore, the proposed scheme is naturally suited to recently available polarization sensitive pulse shapers. These can completely avoid the phase drifts of the interferometer, since both polarization components travel along the same path, and can reach up to kHz modulation rates with high diffraction efficiency \cite{rocha_self-configuring_2025}.

In conclusion, in this work we have demonstrated a proof-of-principle implementation of QSS in a wavelength-multiplexed quantum network. 
The main idea relies on spectral phase modulation performed by the quantum server as supplier of entanglement. 
It achieves independent state modulation for each pair of correlated frequency channels, while the entanglement in another DOF is used to perform and track the security of the protocol. 
To showcase our scheme, we have built an experiment employing only COTS components, including an all-fiber polarization-entangled photon source. 
Our methods extend the capabilities of standard wavelength-multiplexed networks beyond QKD and could pave the way for further advances in quantum cryptography.

\section*{Funding}
This work was financially supported by the Federal Ministry of Education and Research of Germany
(BMBF) through the QuNET initiative.

\section*{Acknowledgements}
The presented results were partially acquired using facilities and devices funded by the Free State of
Thuringia within the application center Quantum Engineering. MCP and CSG are part of the Max Planck
School of Photonics supported by BMBF, Max Planck Society, and Fraunhofer Society.

%

\section*{Appendix: Estimation of asymptotic key rates}
Let us introduce a model to estimate the asymptotic key rates with our system. 
Although the modulation for phase encoding discretizes the temporal modes, we base our analysis on a continuous-wave pump model \cite{neumann_model_2021}. 
We assume that the modulation can be performed at a rate faster than the photon pair production. In this case, every generated photon pair encodes a secret bit of Charlie.

\begin{figure*}
    \centering
    \includegraphics[width=0.975\linewidth]{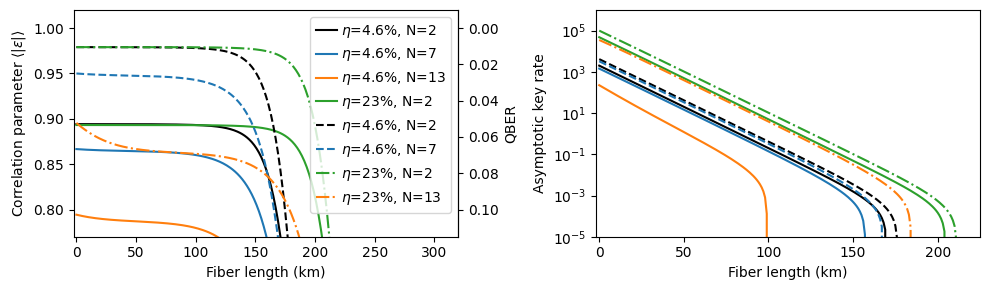}
    \caption{Simulation of the average correlation parameter $\langle|\epsilon|\rangle$ and QBER (left) as well as key rates (right) with fiber distance. Both graphs are calculated with 
    the same 
    source brightness $B=4.7\times10^6$\,pairs/s, 
    allowing for a direct comparison of the effects of visibility $\langle|\epsilon|\rangle$, heralding efficiency $\eta$ and number of users $N$ in the network. 
    Solid lines represent estimations with our experimental visibility $\langle|\epsilon_\text{exp}|\rangle=89.4$\,\%. Dashed lines denote improved visibilities up to $\langle|\epsilon_\text{ideal}|\rangle=98\,\%$, and dash-dot lines represent simultaneously improved visibility and heralding. The black curve specifically represents our experimental conditions. }
    \label{fig:QSS_rates}
\end{figure*}

In general, noise and loss in the communication system reduces the metrics of entanglement and hence the rates in QKD and QSS. The noise is characterized by the QBER, which is affected by two main factors: the measurement error as well as the ratio of true photon pairs to detected single counts. 
The measurement error $e_\text{meas}$ accounts for imperfections of the measurement devices as well as imperfect fidelities of the quantum state. 
Detected single photons $S_j$ at each user $j$ include environmental photons $S_\text{noise}$ (e.\,g. daylight noise), detector dark counts $S_\text{DC}$ and other uncorrelated photons. These photons may arise from pairs that lost their partner during transmission or from uncorrelated frequency channels. The latter contribution is particularly relevant for studies of multiplexed networks \cite{wengerowsky_entanglement-based_2018}. 
The total number of detected single photons for each user are:
\begin{equation}
    S_j = 2S_\text{DC} + S_\text{noise} + (N_\text{MUX}-1)B \eta - (N_\text{MUX}-2)B\eta^2
\end{equation}
Here, we use the source brightness for a given frequency channel $B$ and the heralding efficiency $\eta$ to estimate the single count rate without additional noise. 
The heralding efficiency accounts for coupling loss, loss in the WDMs and other setup components, as well as detector efficiency. Additionally, propagation loss over a distance $L$ can be simulated by multiplying the heralding efficiency with a loss factor $\eta_\text{loss}=e^{-\alpha L}$. For example, the loss coefficient in SMF28 fibers is $\alpha=0.2\,$dB/km at telecom wavelengths. 
Furthermore, if a single detector simultaneously measures all the allocated frequency channels in a fully connected network, the single photon contribution $B\eta$ increases by a factor of $N_\text{MUX}-1$, corresponding to the remaining number of quantum links \cite{wengerowsky_entanglement-based_2018}. Under these conditions, it is relevant to include detector saturation effects as an additional loss factor that depends on the detector dead time $t_\text{dead}$ \cite{lee_large-alphabet_2019,neumann_model_2021}:
\begin{equation}
\eta_\text{sat}=\frac{1}{1 + S_j t_\text{dead}}
\end{equation}

Single photon detections produce probabilistic accidental coincidence events that are proportional to the total single count rates and the selected coincidence window $\tau$, assuming Poissonian distribution \cite{neumann_model_2021}:
\begin{equation}
    R_\text{acc} = S_A \times S_B \times 2\tau
\end{equation}
Since our polarization analyzer has a single detector per user, the accidental coincidences produce an error half of the time. Therefore, the total QBER is obtained from the ratio of error coincidence events with respect to the total coincidence detections as:
\begin{equation}
    Q = \frac{ C_\text{true}  e_\text{meas} + \frac{1}{2}R_\text{acc}}{C_\text{true} + R_\text{acc}}
\end{equation}

For this experiment, the detected brightness is $R_\text{det}=10$\,kpairs/s for the 200\,GHz channel pair at an average heralding ratio of $\sim$ 4.6\,\% and $\sim$80\,\% detection efficiency. This corresponds to a brightness of $B=4.7\times10^6$ pairs/s. The detectors exhibit a deadtime of $t_\text{dead}\sim 40\,$ns, a dark count rate of $S_\text{DC}=500$ counts/s and the total detection jitter amounts to $\sim$50\,ps. Therefore, we use a coincidence window of 100\,ps. 
In the absence of additional loss or noise, i.\,e. 
$S_\text{noise}=0$, the average key rate is $q\times r_\infty\times R_c=2\,$kbits/s, using the factor $q=\frac{1}{2}$ for basis reconciliation.

To contextualize our results in an actual fiber network, let us simulate the increase of QBER with symmetric fiber distance $L$ and number of users $N_\text{MUX}$. The results are displayed in figure \ref{fig:network}. 
Here one can see the impact of increasing the number of measured quantum links $(N_\text{MUX}-1)$ while using a single detector (black, blue and orange solid curves). Essentially, the increased number of uncorrelated single photons increase errors and reduce the key rates.
Indeed, with our current parameters and without optimization of the photon-pair rate, no positive key can be obtained with $N_\text{MUX}>13$ users.

On the other hand, reducing the system loss by 7\,dB from our MZI and other components increases the key rates almost by a factor of 50 (green solid curve). 
In other words, the same key rates are obtained after an additional 35\,km 
of fiber propagation. 
Improving the stabilization of the interferometer has also a strong influence on the visibilities and total key rates (dashed and dash-dot curves). An improvement up to 98\,\% would double the key rates at the same heralding efficiency (black dashed curve). Whereas a simultaneous reduction of 7\,dB loss in our system would increase by $\times100$  the total key rate (green dashed curve). 

Our current experimental system provides an asymptotic key rate of $>100$\,bits/s for a symmetric photon distribution with up to 30\,km of optical fiber (the equivalent of 6\,dB loss per photon), without optimization of photon-pair rate. It also achieves $>10$\,bits/s after $\sim\,$60\,km of fiber (the equivalent of 12\,dB), well within metropolitan distances. 

\bibliography{MyBiblio}

\end{document}